# UY Puppis – A New Anomalous Z Cam Type Dwarf Nova

**Rod Stubbings**

*Tetoora Road Observatory, 2643 Warragul-Korumburra Road, Tetoora Road 3821, Victoria, Australia; stubbo@sympac.com.au*

**Mike Simonsen**

*AAVSO, 49 Bay State Road, Cambridge, MA 0213;mikesimonsen@aavso.org*

**Abstract**

The defining characteristic of Z Cam stars are "standstills" in their light curves. Some Z Cams exhibit atypical behaviour by going into outburst from a standstill. It has previously been suggested that UY Pup had been a Z Cam star, but it was ruled out due to its long-term light curve. However, in December 2015 UY Pup went into outburst and unexpectedly entered into a short standstill instead of returning to quiescence. Furthermore, UY Pup exhibited additional unusual behaviour with two outbursts detected during its standstill. After this standstill UY Pup made a brief excursion to a quiescence state and slowly rose to a longer and well-defined standstill, where it again went into another outburst. Through comparative analysis, research, and observational data of UY Pup, it is evident and thus concluded that it is indeed a Z Cam star, in which renders it to be one of only four known "anomalous Z Cam stars".

## 1. Introduction

Dwarf novae are cataclysmic variable stars with close binary systems in which a cool secondary star fills its Roche lobe and transfers mass to a white dwarf primary (Warner 1995). These systems experience recurring outbursts that last several days or more and range in brightness from 2 to 6 magnitudes in V with intervals that can vary from days to decades.

Dwarf novae can be classified into three subtypes: U Gem-type stars that have repeated outburst cycles, SU UMa-type stars that show both short and very long outbursts (superoutbursts), and Z Cam stars. The Z Cam subtype, a rare class of dwarf novae, is characterised by standstills at an intermediate brightness level below the outburst maximum and above the quiescence state. Currently there are only 22 known Z Cam stars (Stubbings & Simonsen 2014).

A standstill typically starts at the end of an outburst and remains at a relatively constant brightness of 1 to 1.5 magnitudes below maximum light. These standstills have been known to last from several days to several months. To date the orbital periods of Z Cam stars range from 3 hours to over 8.4 hours (Simonsen 2014). The observational data imply that a bona fide standstill will last up to the length of the mean outburst cycle or longer.

Simonsen (2009) coordinated an observing campaign, the Z CamPaign, to monitor their long-term light curves and correctly classify Z Cam systems. Another aim of the campaign was to determine if any Z Cam stars enter into outburst from a standstill (Simonsen 2010). The results thus far have shown that nine Z Cam stars go back into outburst from a standstill: Z Cam, HX Peg, AH Her, HL CMa, UZ Ser, AT Cnc, Leo5, V513 Cas and IW And (Simonsen *et al.* 2014).

Two systems, V513 Cas and IW And, displayed unusual behaviours in their light curves by having brief excursions to quiescence in addition to going into outburst after a standstill (Simonsen 2011). The light curve of V513 Cas is shown in **Figure 1**.

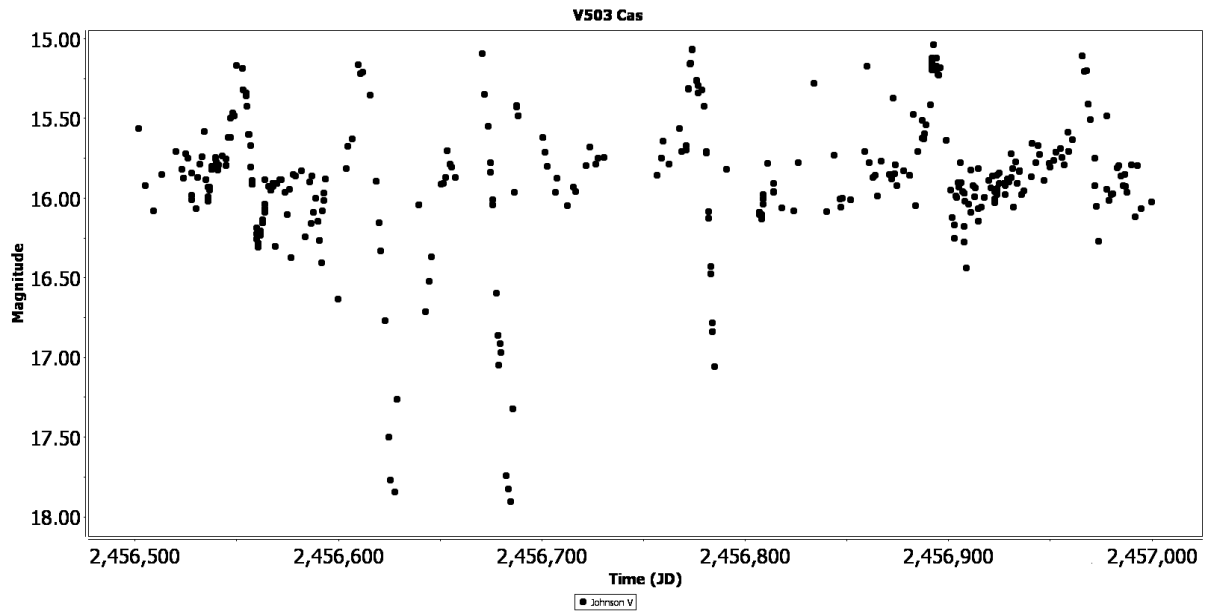

**Figure 1.** The AAVSO light curve of V513 Cas from JD 2456500 - 2457000 (July 2013 – December 2014). The observed pattern of standstills and outbursts can be clearly seen as well as brief excursions to quiescence. Black dots are Johnson V.

The two anomalous Z Cam stars V513 Cas and IW And were studied further by Szkody (*et al.* 2013) to understand the reason why outbursts occur after standstills. The data revealed no explanation as to why outbursts occur from standstills instead of returning to the quiescence state, which is typical behaviour for Z Cam systems (Szkody *et al.* 2013). This phenomenon on V513 Cas and IW And was further investigated by Hameury and Lasota (2014), who concluded that it is possibly due to magnetic activity of the secondary star which might be considerably stronger than in other systems. As shown in **Figure 2**, the recently classified anomalous Z Cam star ST Cha displays the same remarkable behaviour as V513 Cas (Simonsen *et al.* 2014). ST Cha spends most of the time in standstills followed by outbursts, brief excursions to quiescence, and in standstills emerging from quiescence.

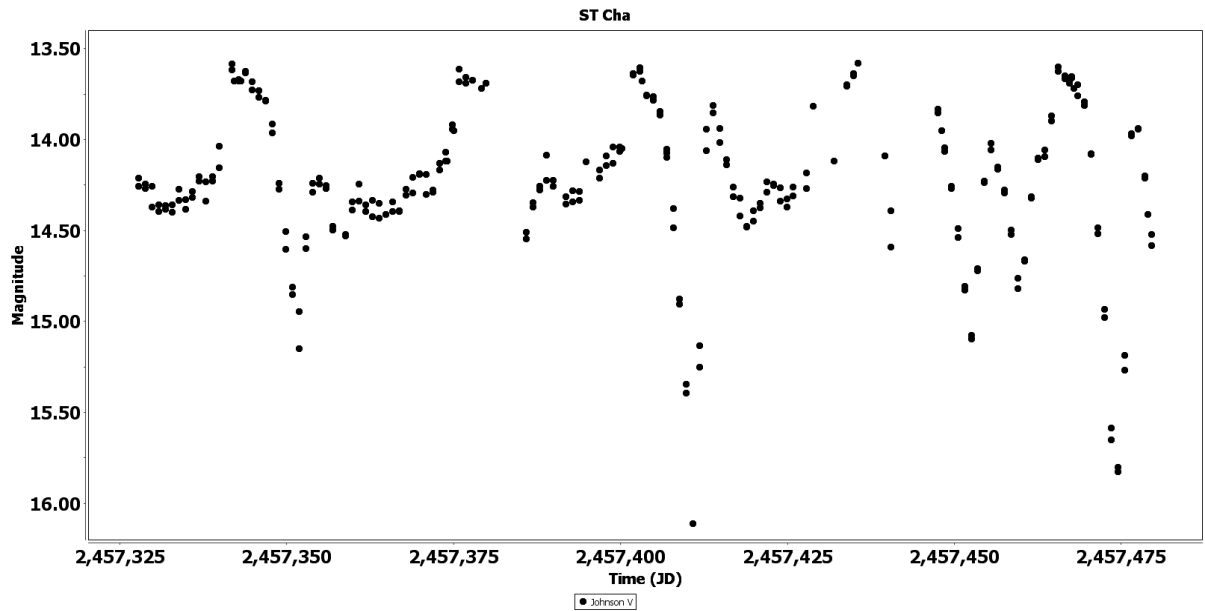

**Figure 2.** The AAVSO light curve of ST Cha from JD 2457320 - 2457480 (October 2013 - April 2016) featuring standstills followed by outbursts and brief minima. Black dots are Johnson V.

UY Pup has shown numerous shared features in its light curve with anomalous Z Cam stars, in particular ST Cha. The commonalities between these stars and UY Pup include frequent outbursts, followed by a number of small duration standstills for an extended period of time, brief excursions to quiescence, and standstills starting from quiescence.

## 2. History

A spectroscopic study of UY Pup in outburst showed it to be a low inclination dwarf novae and an estimated orbital period of 10.20 hours (Lockley *et al*. 1999). John R. Thorstensen *et al.* (2004) found a more accurate period of 11.50 hours. Bateson (1991) analysed the first 90 outbursts of UY Pup and found the mean cycle to be ~34.93 days. Long and short outbursts occurred, with the long outbursts being more frequent and both types of outbursts showing little difference in the mean brightness (Bateson 1991). There was no mention of UY Pup as a possible Z Cam star from the data collected by Bateson. UY Pup was measured in outburst at V = 14.1 with B-V = 0.09 and U-B = -0.78 (Bruch and Engel 1994), who classified UY Pup as a possible Z Cam type star.

**3. Observations**

UY Pup has an observing window from October through June and consequently has large gaps in the recorded observations. UY Pup has been monitored by the Royal Astronomical Society of New Zealand (RASNZ) and AAVSO in which the brightness varies from magnitude ~13.0 in outburst to ~16.1 V in quiescence. Furthermore, UY Pup has been monitored continually by one of the authors of this paper (RS) from the year 1995 to present. The outburst and quiescence pattern had become familiar over this time frame, especially in the later years due to closer monitoring. To determine the mean cycle of UY Pup during this period, consecutive outbursts with intervals of less than 60 days were selected. The conclusion from this analysis was that there were 67 outbursts which were considered consecutive and led to a mean cycle of ~31.86 days.

The increased monitoring of UY Pup on its rise and decline detected an interesting result that revealed some of the short outbursts were in fact part of the long outbursts. The first evidence of this was noticed in December 2015 around JD 2457365 when UY Pup was near minimum at a visual magnitude of 15.6 (Figure 3). Five days later it rose to 14.4 and varied by ~ 0.2 magnitude for 11 days, then went into outburst on JD 2467386, reaching a visual magnitude of 13.8. The fall from this outburst lasted about 7 days, and then UY Pup stalled near magnitude 14.5 for 2 weeks, before going back into outburst on JD 2457417, where it attained a visual magnitude of 13.7. UY Pup declined from this outburst for about 4 days before stalling once again near visual magnitude 14.5 for a further 30 days, and then fading to quiescence and varying in magnitude between ~15.6 and ~15.9 in V.

This small standstill started on JD 2457370 and lasted for around 66 days; two outbursts were observed in this time frame. After the brief excursion to quiescence at ~15.9 V UY Pup started to slowly rise to a longer and well-defined standstill, evolving from the quiescence state on JD 2457474 and continuing for about 71 days. The magnitude varied ~0.5 during this period before going into an outburst at V = 13.4. **Figure 3.**

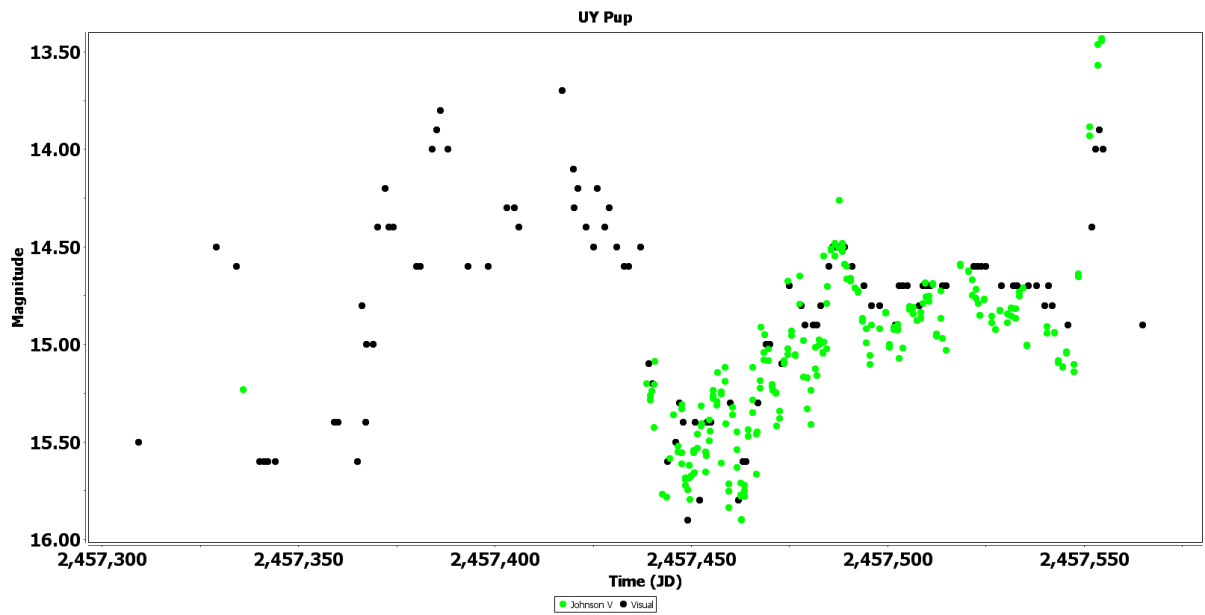

**Figure 3.** AAVSO light curve showing the first well-defined standstill of UY Pup beginning on JD 2457474 - 2457545 (March 26, 2016 - June 5, 2016 ) and lasting 71 days before going into outburst. Black dots are visual data, green dots are Johnson V.

A comparative analysis of the light curve of ST Cha from JD 2457336 to 2457392 was undertaken, in which showed a very similar light curve. Furthermore, the anomalous features in the light curves of both UY Pup and ST Cha are also evident in the standstills that have emerged from quiescence in ST Cha. **Figure 4**.

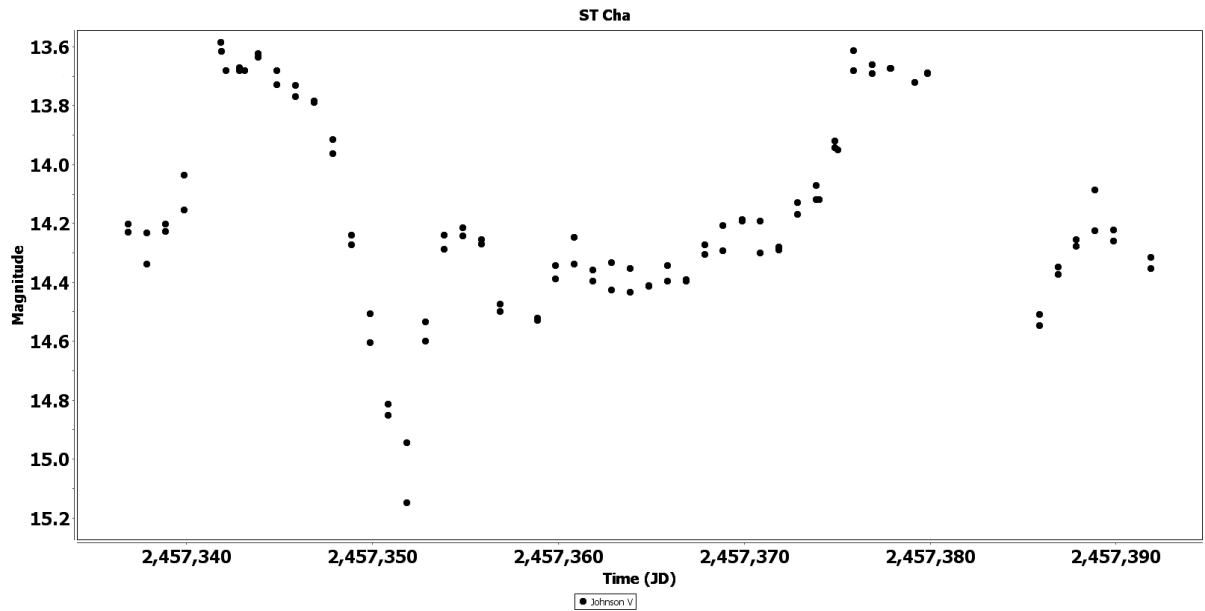

**Figure 4.** AAVSO light curve of ST Cha from JD 2457336 - 2457392 (November 2015 – January 2016) showing a standstill starting from the quiescence state. Black dots are Johnson V.

Outbursts from standstills are quite rare in Z Cam stars, so it was decided that previous data on UY Pup should be examined. As a result of further research another well-defined anomalous standstill in October 1996 (starting on JD 2450389) was found **Figure 5.** The total duration of this standstill was around 228 days and included two brief fades. The visual magnitudes of the fades observed were 15.0 and 15.4, which lasted a few days before going straight back into outburst on each occasion. In this time frame, UY Pup went into outburst five times from standstills in the low to mid 14$^{th}$ magnitude range.

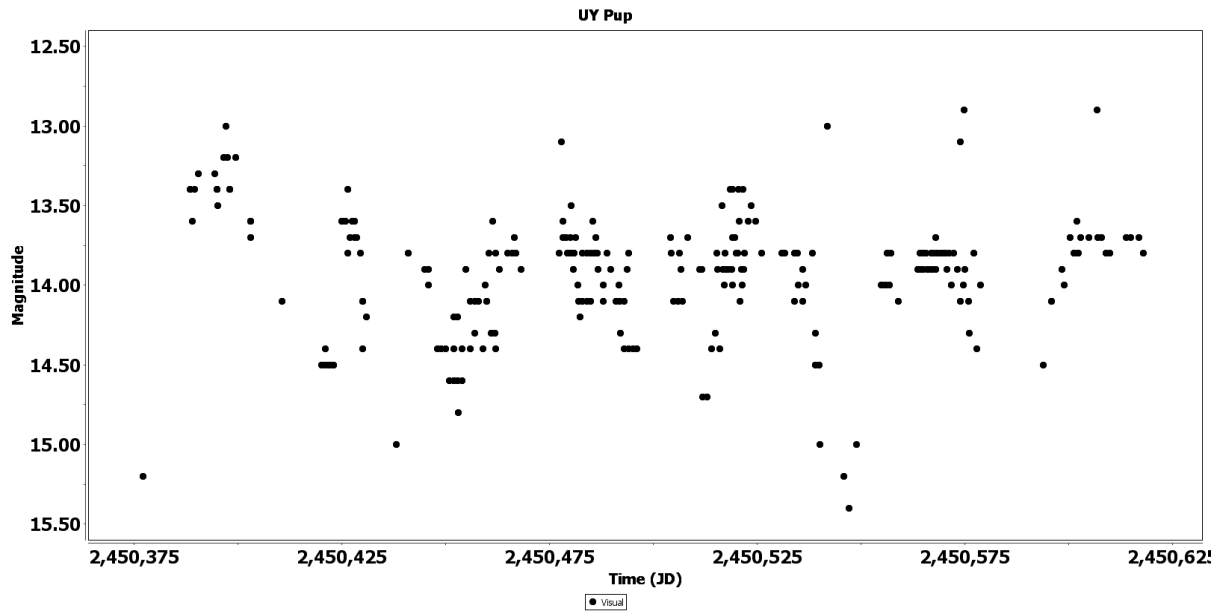

**Figure 5.** Hidden in the long-term AAVSO light curve, the almost continuous standstill of UY Pup followed by outbursts can be seen. Beginning on JD 2450389 (October 1996) the system shows a pattern of standstills in the mid-range of ~13.7 to ~14.3 and outbursts lasting for 228 days to June 1997 (JD 2450617). Black dots are visual data.

## 4. Conclusion

UY Pup has been subject to more thorough monitoring since 1995 and as a result several features and new findings that are atypical behaviours in this particular star have been highlighted. It is now evident that part of the long outbursts are associated with the short outbursts and on some occasions the short outbursts fail to decline to quiescence and enter into small standstills.

UY Pup shows definitive signs of being one of the anomalous Z Cams stars like IW And, V513 Cas, and ST Cha. Standstills occur on the decline from outburst as seen in typical Z Cam systems, as well as brief fades to quiescence. The data also show that standstills can occur in the rising branch from the quiescence state and the emergence of outbursts following these standstills, as seen the light curve of ST Cha in **Figure 2.** UY Pup has one of the longest orbital periods for a Z Cam star at 11.5 hours, which makes it an interesting object for further study.

Past observations have presumed that all typical Z Cam standstills were terminated by the decline to quiescence. Since the introduction of the Z CamPaign and closer monitoring of these systems it is evident that some Z Cam stars have more complex behaviours in their light curves. In addition to the outbursts from standstills the light curves show brief fades to quiescence and standstills starting from the quiescence state. As a result of the research and data it has been concluded that there are now four anomalous Z Cams stars: IW And, V513 Cas, ST Cha and UY Pup.

**5. Acknowledgements**

We acknowledge with thanks the variable star observations from the AAVSO International Database (Kafka 2016) contributed by observers worldwide and used in this research. This research has made use of the International Variable Star Index (VSX) database (Watson *et al.* 2015), operated at AAVSO, Cambridge, Massachusetts, USA. We would also like to thank the anonymous referee whose comments and suggestions were helpful.

**References**

Bateson, F.M. 1991, *Publ. Var. Star Sect. Roy. Astron. Soc. New Zealand,* 16, 80.

Bruch, A., and Engel, A. 1994, *Astron. Astrophys., Suppl. Ser.,* 104, 79.

Hameury, J. M., and Lasota, J. P. 2014, *Astron. Astrophys., 569*, A48

Kafka, S. 2016, observations from the AAVSO International Database <http://www.aavso.org>.

Lockley, J. J., Wood, J. H., Jones, D. H. P., and Mineshige, S. 1999, *Astrophys. Space Sci.,* 266, 453.

Simonsen, M. 2009, *The Z Cam List*, <https://sites.google.com/site/thezcamlist/the-list>.

Simonsen, M. 2010, in *The Society for Astronomical Sciences 29th Annual Symposium on Telescope Science*, Society for Astronomical Sciences, Rancho Cucamonga, CA, 87.

Simonsen, M. 2011, *J. Amer. Assoc. Var. Star Obs.,* 39, 66.

Simonsen, M. 2014, *in The Society for Astronomical Sciences 33rd Annual Symposium on Telescope Science*, Society for Astronomical Sciences, Rancho Cucamonga, CA, 143.

Simonsen, M., Bohlsen, T., Hambsch, J., and Stubbings, R. 2014, *J. Amer. Assoc. Var. Star Obs.*, 42, 199.

Simonsen, M*., et al.* 2014, *J. Amer. Assoc. Var. Star Obs.*, 42 177.

Stubbings, R., and Simonsen, M. 2014, *J. Amer. Assoc. Var. Star Obs.,* 42, 204.

Szkody, P., *et al.* 2013, *Publ. Astron. Soc. Pacific*, 125, 1421.

Thorstensen, J. R., Fenton, W. H., and Taylor, C. J. 2004, *Publ. Astron. Soc. Pacific*, 116, 300.

Warner, B. 1995, *Cataclysmic Variable Stars*, Cambridge Univ. Press., Cambridge.

Watson, C., Henden, A. A., and Price, C. A. 2015, *AAVSO International Variable Star Index VSX* (Watson+, 2006–2015; http://www.aavso.org/vsx).